\documentclass[twocolumn,prb, showpacs]{revtex4}
\usepackage{graphicx}
\usepackage{dcolumn}
\usepackage{bm}
\begin{document}

\title{Electrochemical De-intercalation, Oxygen Non-stoichiometry,\\ and
Crystal Growth of Na$_x$CoO$_{2-\delta}$}

\author{F. C. Chou$^1$}
\author{E. T. Abel$^{1,2}$}
\author{J. H. Cho$^{2}$}
\altaffiliation[On leave ]{from Physics Department, Pusan National
University, Korea.}
\author{Y. S. Lee$^{1,2}$}
\affiliation{%
$^1$Center for Materials Science and Engineering,
Massachusetts Institute of Technology, Cambridge, MA 02139}%
\affiliation{%
$^2$Department of Physics,
Massachusetts Institute of Technology, Cambridge, MA 02139}%

\date{\today}

\begin{abstract}
We report a detailed study of de-intercalation of Na from the
compound Na$_{x}$CoO$_{2-\delta}$ using an electrochemical
technique.  We find evidence for stable phases with Na contents
near the fractions $x\simeq$1/3, 1/2, 5/8, 2/3, and 3/4. Details
regarding the floating-zone crystal growth of Na$_{0.75}$CoO$_2$
single crystals are discussed as well as results from magnetic
susceptibility measurements.  We observe the presence of
significant oxygen deficiencies in powder samples of
Na$_{0.75}$CoO$_{2-\delta}$ prepared in air, but not in single
crystal samples prepared in an oxygen atmosphere.  The oxygen
deficiencies in a Na$_{0.75}$CoO$_{2-\delta}$ sample with $\delta
\sim$ 0.08 remain even after electrochemically de-intercalating to
Na$_{0.3}$CoO$_{2-\delta}$.\\
\end{abstract}

\pacs{74.62.Bf, 74.10.+v, 74.62.Dh, 61.10.Nz}

\maketitle

\section{\label{sec:level1}Introduction\protect\\ }

Following the discovery of Li$_x$CoO$_2$ as a good cathode
material for use in solid state batteries,~\cite{Mizushima1980}
the Na$_{x}$CoO$_2$ compound was also identified as a candidate
having high energy density with good
reversibility.~\cite{Molenda1983} Further research on
Na$_{x}$CoO$_2$ was spurred by the observation of its large
thermoelectric power.~\cite{Terasaki1997} The discovery of
superconductivity in hydrated Na$_{0.3}$CoO$_2$.1.3H$_2$O with
T$_c$ $\sim$ 4.5 K has led to much recent interest in this
material.~\cite{Takada2003}  There is the intriguing possibility
that strong electronic correlations play an important role in the
superconductivity.  For further experimental work, high quality
single crystals with precisely controlled Na contents are greatly
desired.  The traditional flux method of crystal growth, using
NaCl as a flux, typically yields very thin crystals of limited
size.~\cite{Fujita2001} Recently, several groups have reported
growing large crystals of Na$_{x}$CoO$_2$ using the optical
floating-zone technique with a stoichiometric self
flux.~\cite{Jin2003,Chen2004,Chou2004,Prakhakaran2003} For these
crystals, different methods have been used to control the Na
content $x$, and the reported measurements of the physical
properties show a larger degree of variability compared with
measurements on powder samples.

Chemical de-intercalation using Br$_2$ has been widely utilized to
de-intercalate Na ions from Na$_{x}$CoO$_2$.  By adjusting the Br
concentration of the solution, various phases with different $x$
can be obtained.~\cite{Foo2004}  We have developed an
electrochemical technique for de-intercalating Na.  The sample is
immersed, along with a counter electrode and a reference
electrode, in a solution of NaOH.  By applying a voltage between
the sample and counter electrode, an electrochemical double layer
($\sim$10$\AA$ in thickness) forms at the sample's surface.  The
high gradient of reactant together with the applied over-potential
induces the desired chemical reaction. Using this method, we can
monitor the integrated charge during the de-intercalation process.
Moreover, the charging rate for the sample can be readily
controlled, and thus it may be easier to achieve equilibrium
phases. Using this electrochemical technique, we have employed a
potential step method to extract information on the stable phases
of Na$_{x}$CoO$_2$.

The possibility of oxygen deficiency in Na$_{x}$CoO$_2$ is an
important issue, especially in determining the valence of the Co
ions. Significant oxygen deficiency is generated in the related
Li$_x$CoO$_{2-\delta}$ system when x is below $\sim$0.5,
~\cite{Venkatraman2002} which prohibits further changes to the
valence of Co above +3.5.  The instability of the Co$^{4+}$
valence state may also make Na$_{x}$CoO$_2$ prone to be oxygen
deficient, especially for $x$ below $\sim$0.5. In an earlier
study, Molenda $\it{et ~al.}$ found that
Na$_{0.7}$CoO$_{2-\delta}$ prepared with an oxygen partial
pressure of $\sim$0.2 atm had an oxygen defect level as high as
0.073.~\cite{Molenda1989}  The degree of oxygen deficiency will
likely depend strongly on the method of sample preparation. We
have investigated the presence of such deficiencies in both powder
and single crystal samples. In addition, we have measured how the
de-intercalation process affects the deficiency level. Clearly, if
oxygen deficiencies exist, they would strongly affect the
electronic structure of the CoO$_2$ planes.

This paper is organized as follows: Section II discusses the
sample preparation and measurement methods used.  Our results,
which are intimately connected to our sample preparation methods,
are also presented in this section. These results include data
from thermal gravimetric analysis, x-ray diffraction, magnetic
susceptibility experiments, and measurements of the integrated
charge passed through the electrochemical cell. Section III
contains a discussion and conclusions.

\section{\label{sec:level1}Experimental Details and Results\protect\\}

The single crystals used in this study were grown using the
travelling-solvent floating-zone (FZ) technique with a
stoichiometric self-flux.  The feed rod was melted and
re-crystallized using a CSI four-mirror furnace with Xe-lamps
under various pulling rates.~\cite{Chou2004} Samples with various
Na contents were prepared by electrochemically de-intercalating an
as-grown Na$_{0.75}$CoO$_2$ crystal.  The basic set up for the
electrochemical cell was described in our previous
work.~\cite{Chou2004}  Crystals with x = 0.67, 0.5 and 0.3 were
prepared by ramping the applied voltage in small steps (0.001
mV/step) from the initial open circuit potential to final constant
potentials of $\sim$ 0 V, 0.3 V, and 0.7 V (relative to a Ag/AgCl
reference) respectively.  The samples were determined to have a
single structural phase with the expected c-axis lattice constant
reported in the literature by  x-ray diffraction, and the Na
content was determined by electron probe microanalysis (EPMA). The
magnetic properties were characterized using a SQUID magnetometer
(Quantum Design MPMS-XL) with an applied field of 100 Oe and 1
Tesla. The oxygen and water contents were calculated from the
weight loss/gain of samples in an oxygen atmosphere using thermal
gravitimetric analysis (Perkin-Elmer TGA7), and the melting points
were checked by differential thermal analysis (Perkin-Elmer DTA7).

\subsubsection{\label{sec:level2}Floating zone crystal growth and characterization\\}

\begin{figure}
\includegraphics[width=3.5in]{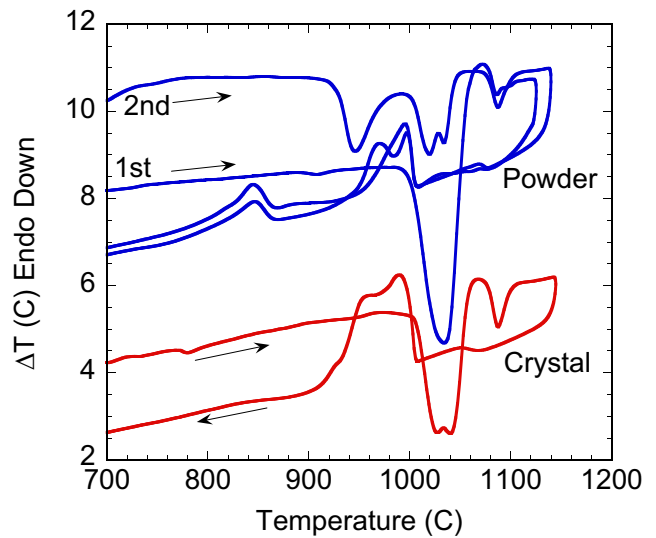}
\caption{\label{fig:EC-eps1} (color online) DTA scan for
Na$_{0.75}$CoO$_2$ powder and FZ crystal.  The scan has been done
with warming/cooling rate of 40 C/min in an oxygen environment.}
\end{figure}

In order to test whether a direct re-crystallization process is
suitable for the crystal growth of Na$_{0.75}$CoO$_2$ ,
differential thermal analysis (DTA) measurements on stoichiometric
Na$_{0.75}$CoO$_2$ powder were performed.  The results are shown
in Fig.~\ref{fig:EC-eps1}. The sample melts incongruently near
$\sim$1020C and subsequently enters a completely liquid phase near
$\sim$1075C. The 1020C peak is observed to have a doublet
structure which suggests that the incongruent melting process
crosses two tie lines upon heating, each associated with a solid
phase of different stoichiometry.  It is possible that the
solidification process during the crystal growth may only quench
one of the solid phases. Since the temperature difference between
these two solid phases is small ($\sim 15-20$ C), a mixture of two
solid phases is very likely.  As a result, large single-phase
crystals are difficult to produce.  This difficulty is exacerbated
by the conditions of continuous Na loss during the growth.  These
difficulties may underlie the discrepancies between powder and FZ
crystal measurements, and even between measurements performed on
FZ crystals grown by slightly different procedures.~\cite{
Chen2004, Prakhakaran2003, Bayrakci2003, Sales2004}

To address Na vapor loss, a rapid-heating technique has been
developed for powder preparation and has proven to be reliable in
obtaining single phase samples between x = 0.65 and
0.75.~\cite{Motohashi2001} This minimizes the amount of a possible
CoO impurity phase.  The presence of a CoO impurity phase can be
detected by comparing DTA scans of Co$_3$O$_4$ (not shown) and
powder Na$_{0.75}$CoO$_2$.  As shown in Fig.~\ref{fig:EC-eps1},
the powder Na$_{0.75}$CoO$_2$ shows a pronounced solidification
exothermic peak $\sim$860C, which corresponds to the CoO phase
formation. On the other hand, the single crystal lacks
significant signature of a CoO impurity phase.  The CoO phase does
not necessarily become solidified in the crystal if Na loss is
constantly compensated. Since it is difficult to maintain a
constant flux ratio, it is preferable to fix the heating power and
adjust the feeding rate of the stoichiometric feed rod in order to
maintain a stable molten zone.

We find that the FZ crystals have a morphology which depends
significantly on the pulling rate.  The c-axis is always
perpendicular to the growth direction and the crystal is
relatively easy to cleave along the ab plane.  A slow pulling rate
below 2 mm/h yields crystals with many misaligned grains,
independent of the quality of the seed. The misaligned grains of
1-3 mm size are hard to cleave, and the grain boundaries contain a
CoO impurity phase.  On the other hand, fast pulling rates (higher
than 8 mm/h) can usually prevent misalignment along the growth
direction and yield large rod-shaped single crystals.  The
crystals grown via fast pulling are easy to cleave into large flat
pieces.  For this pulling rate, a low level of CoO impurity
inclusion can be achieved with proper adjustment of extra feed to
maintain a stable molten zone, i.e. when the correct flux ratio is
obtained so that solid phase Na$_{0.75}$CoO$_2$ precipitates from
the flux continuously.

The visual appearance of the surface of the crystalline rod is not
a reliable indicator of overall crystal quality.  Crystals grown
with an intermediate pulling rate of 4 mm/hr have an outer surface
with a mirror-like finish, however the core region of the rod is
composed of small misaligned domains.  There appears to be a high
level of CoO impurity phase within the grain boundaries inside the
core region.  EPMA results indicate the Na content to be near 0.75
to within the experimental error ($\pm0.08$ from different batches
and regions) and there is no significant difference of the Na
concentration between the core and outer ring regions of the
crystalline rod.

\begin{figure}
\includegraphics[width=3.5in]{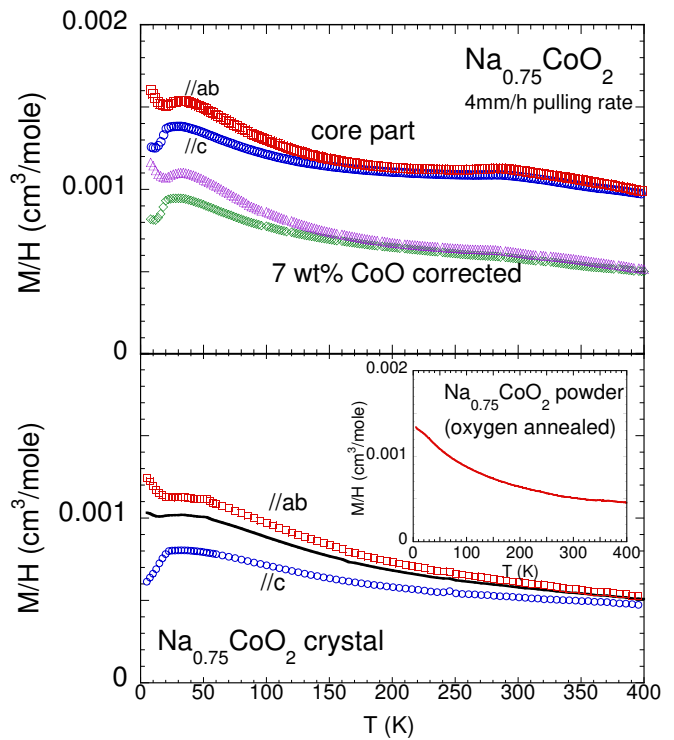}
\caption{\label{fig:EC-eps2} (color online) Magnetic
susceptibilities of a floating zone Na$_{0.75}$CoO$_2$ crystal,
which has been grown with 4 mm/hr pulling rate.  The upper panel
shows data from the core part, with and without $\sim$7 wt\%\ CoO
correction.  The lower panel is the outer ring part without 290K
peak and the solid line corresponds to its powder average.  Inset
shows measurement for a powder sample with x = 0.75}
\end{figure}

In much of the literature there are subtle discrepancies between
magnetic susceptibility measurement performed on FZ single
crystals of Na$_{x}$CoO$_2$ (x $>$0.65) grown by slightly
different methods.  The primary discrepancies regard the magnitude
of the susceptibility and the presence of a broad feature near 290
K.  We mechanically separated the core material from the outer
shiny material from our crystal grown at the 4mm/hr pulling rate.
We then measured both materials separately in a SQUID
magnetometer.  The top panel of figure 2 shows the susceptibility
of the core region.  Since CoO has an antiferromagnetic transition
near 290 K, it is reasonable to assume that the broad feature near
290K is due to the presence of a CoO impurity phase. After
subtracting a 7 wt\%\ CoO contribution, the magnitude and shape of
the curves better agree with measured powders.  The bottom panel
of Fig.~\ref{fig:EC-eps2} shows the susceptibility measured on the
shiny outer region along the c- and ab-directions. When a powder
average is taken of the two directions, a curve strongly
resembling the powder susceptibility is recovered.  Hence, the
shiny outer regions of the as-grown crystal are largely free of
CoO impurities.  These data also illustrate that the magnetic
susceptibility is quite sensitive to CoO content; even if no CoO
was detected using x-ray diffraction, it may be noticeable in the
susceptibility.

\subsubsection{\label{sec:level2}Electrochemical de-intercalation\\}

The electrochemical cell was prepared as described
previously,~\cite{Chou2004} with Na$_{0.75}$CoO$_{1.92}$ powder as
the working electrode, platinum as the anode, Ag/AgCl as the
reference electrode, and 1M NaOH as the electrolyte.  We have
employed a potential step method to extract phase information on
Na$_{x}$CoO$_2$, similar to that used previously to study to
La$_2$NiO$_{4+\delta}$ and related compounds.~\cite{Bhavaraju1996}
In the potential step method, the over-potential at the sample is
incremented in small steps.  After each step, the potential is
held constant for a sufficiently long time such that the current
decays to a nearly constant background level, indicating that
equilibrium is nearly achieved.  Hence, the reference potential
measured for the cell can be related to the chemical potential of
Na in the sample.  Since extremely long times may be required to
achieve equilibrium, in practice the voltage is held constant
until the current falls below a threshold value.  At this point,
the next voltage increment is added.  This procedure ensures
consistent charge dynamics at each new voltage.

\begin{figure}
\includegraphics[width=3.5in]{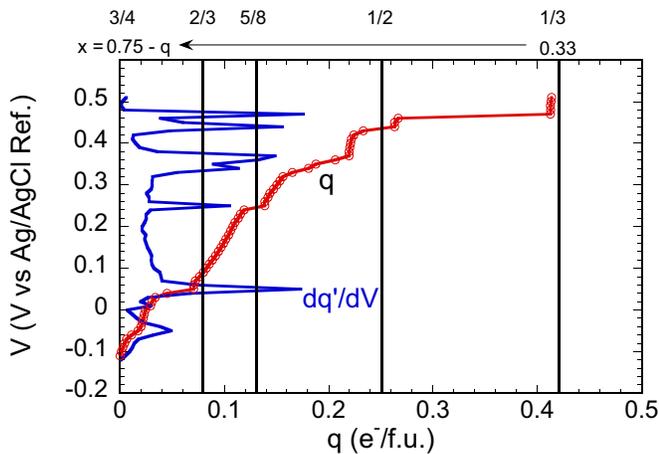}
\caption{\label{fig:EC-eps3} (color online)Applied voltage versus
accumulated charge which passed through the electrochemical cell,
where powder sample of x = 0.75 is used as working electrode.}
\end{figure}

In our experiments, the voltage was stepped from the initial open
circuit potential of about -0.15 V to 0.6 V in 5 mV steps
(measured in reference to Ag/AgCl).  After each voltage step, the
current through the electrolyte was allowed to decay to 0.25 of
the initial current. The current was then extrapolated to infinite
charging time assuming exponential decay, and then integrated to
obtain $dq$ for each new voltage increment (where $dq$ is the
extrapolated charge which passes through the electrodes as
equilibrium is approached). The total accumulated charge for all
voltage steps, $q$, (in electrons per formula unit of NaCoO$_2$)
is plotted for each value of the applied voltage $V$, as shown in
Fig.~\ref{fig:EC-eps3}. Also in the figure, we plot
$dq^\prime/dV$, where $dV$ is the 5 mV voltage increment and
$dq^\prime$ is the actual integrated charge during the time the
voltage was held constant.  (Hence, $dq^\prime/dV$ is
qualitatively similar to $dq/dV$, except the magnitude is
smaller.)  The plateau regions (where $V$ is roughly constant for
a range of $q$) indicate the coexistence of two phases.  Thus, the
high-$q$ side of each plateau region indicates a Na content
corresponding to a single stable phase.  The voltages at which
these stable phases occur are marked by the peaks in
$dq^\prime/dV$.

The proposed half reaction at the working electrode is\\

$Na_{0.75}CoO_{1.92} + q (OH^{-}) \rightarrow
Na_{0.75-q}CoO_{1.92} + q e^{-} + q (NaOH).$\\

\noindent Since the initial powder sample is x = 0.75, we can
identify stable phases with $x \simeq$ 2/3, 5/8, 1/2, and 1/3 near
the end of each plateau region. We note that the two plateau
regions between $x \sim$ 5/8 and 1/2 suggest that there are two
narrowly separated phases of close stoichiometry (within $\sim$
0.02) for this Na content. In addition, the chemical potential
difference between $x=$ 1/2 and 1/3 is extremely small and this
two-phase region is the widest among the observed stable phases.
The close proximity of the chemical potential between the phases
with x = 1/2 and x = 1/3 indicates that they are close in free
energy, and thus may have similar physical properties.  This is
consistent with the observation that the susceptibilities and Co
Knight shift data above $\sim$100K for both phases are quite
similar. ~\cite{Chou-mag, Imai-unpub, Foo2004}

\begin{figure}
\includegraphics[width=3.5in]{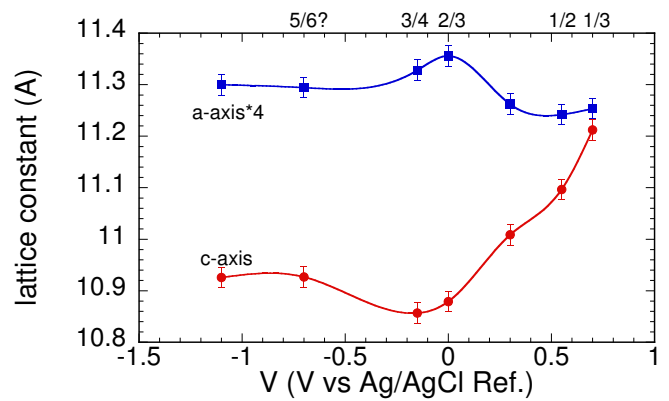}
\caption{\label{fig:EC-eps4} (color online) Ex-situ x-ray
diffraction c-axis versus applied voltage.  The applied voltage
has been ramped up (0.001 mV/s) from the initial open circuit
potential to the final equilibrium potential and stay at the final
potential for one day each.}
\end{figure}

Fig.~\ref{fig:EC-eps4} shows the results of an ex-situ x-ray
diffraction measurements on samples prepared with various applied
electrochemical potentials. Using Na$_{0.75}$CoO$_2$ powder as the
starting material, we applied a potential which was ramped from
the initial open circuit potential to a final value at a rate of
0.001 mV/s. The voltage was then held at the final potential for
24 hours, after which the sample was removed from the cell and
immediately placed into the x-ray diffractometer.  The samples
corresponding to the end points at 0.7V and -1.2V were allowed to
equilibrate for one week and one month respectively to ensure
total conversion.  We find that the c-axis lattice constant
increases from $\sim$ 10.8 $\AA$ to $\sim$11.2$\AA$ with
increasing potential, which agrees fairly well with the results on
Br de-intercalated samples for x = 0.75 and x = 0.3
respectively.~\cite{Foo2004}  Samples prepared with a negative
over-potential ($<$~$-0.5 V$) will have Na ions intercalated back
into the starting compound Na$_{0.75}$CoO$_2$.  Hence, phases with
higher $x$ can be formed, and it appears that one such phase has a
stoichiometry with $x$ near 5/6.

\subsubsection{\label{sec:level2}Oxygen non-stoichiometry\\}

\begin{figure}
\includegraphics[width=3.5in]{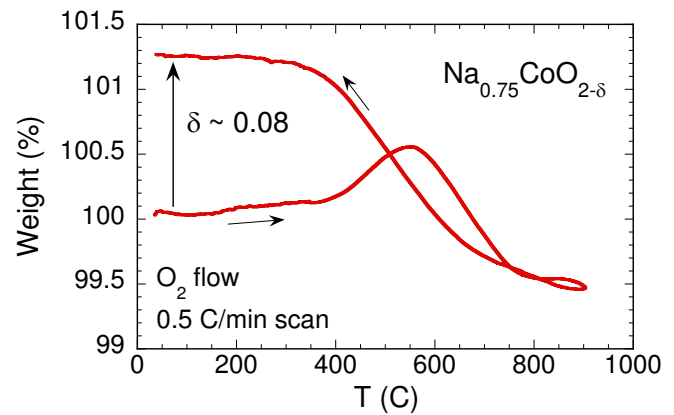}
\caption{\label{fig:EC-eps5}  (color online) The weight versus
temperature of a powder sample with x = 0.75, which was quenched
from 900C to room temperature in the air.}
\end{figure}

It has been reported that significant oxygen deficiencies are
generated in the related Li$_x$CoO$_{2-\delta}$ compound when x is
below $\sim$0.5,~\cite{Venkatraman2002} impeding higher valence
changes to the Co ions.  It is therefore likely that
Na$_{x}$CoO$_2$ is also prone to be oxygen deficient, especially
for x below $\sim$0.5.  This may prevent the valence of the Co
ions to be raised much beyond Co$^{+3.5}$.  In fact,
Na$_{0.7}$CoO$_{2-\delta}$ prepared with an oxygen partial
pressure of $\sim$0.2 atm has been reported to have an oxygen
defect level as high as 0.073.~\cite{Molenda1989} By performing
TGA measurements in an O$_2$ environment, we find that powder
samples of Na$_{0.75}$CoO$_{2-\delta}$ (prepared at 900C and
quenched in air) have a value of $\delta$ very close to
$\sim$0.08, as shown in Fig.~\ref{fig:EC-eps5}.  In contrast, we
find that a FZ crystal of Na$_{0.75}$CoO$_{2}$ (quenched from high
temperature in an oxygen atmosphere) does not pick up oxygen under
the same annealing process.  Hence, for the as-grown single
crystals with $x=0.75$, the oxygen deficiency level is small
($\delta \sim 0$).

For the powder samples of Na$_{0.75}$CoO$_{1.92}$, the oxygen
non-stoichiometry remains even after the electrochemical
de-intercalation process.  Such a sample was used in the potential
step experiment shown in Fig.~\ref{fig:EC-eps3}.  To reduce the Na
content from 0.75 to 0.33, a total charge of 0.42 e$^-$/f.u. is
required, which is indicated by the position of steepest slope
near $\sim$0.42 e$^-$/f.u. as shown in the figure. If the oxygen
deficiencies are being compensated, then an additional charge
equal to 0.16 e$^-$/f.u. would be required through the reaction\\

$Na_{0.33}CoO_{2-\delta}$  + $2\delta(OH)^{-}$ $\rightarrow$
$Na_{0.33}CoO_{2}$ + $2\delta e^{-}$ + $\delta(H_{2}O).$\\

\noindent The fact that no additional charge is observed during
the electrochemical charging suggests that the oxygen deficient
state, Na$_{0.33}$CoO$_{1.92}$, is the final equilibrated phase.
We find that upon cooling Na$_{0.33}$CoO$_{1.92}$ from 900C in
O$_2$, the sample does not pick up additional oxygen to yield
Na$_{0.3}$CoO$_2$, in contrast to the $x=0.75$ compound.  By means
of titration experiments, Karppinen $\it{et~al.}$ have shown that
the Co valence of fully hydrated, partially dehydrated, and
non-hydrated Na$_{0.36}$CoO$_2$ is essentially identical in all
three compounds with a value of +3.48.~\cite{Karppinen2004} Hence,
it appears that the oxygen deficiency level of $\delta \sim 0.08$
keeps the Co valence state near +3.5, even for Na contents
significantly less that $x=0.5$.

\section{\label{sec:level1}Discussion and Conclusions\protect\\}

The various stable phases that we have observed upon
electrochemical de-intercalation may be related to preferred Na
concentrations for the formation of ordered Na superlattices.
Various types of Na superstructures in Na$_{x}$CoO$_2$ have been
identified in both experimental and theoretical
work.~\cite{Shi2004,Baskaran2003,Huang2004,Zandbergen2004}  For a
structure with hexagonal symmetry, the systematic loss of one out
of six Na ions from the lattice naturally describes a series of
samples with $x$ near 5/6=0.83, 4/6=0.67, 3/6=0.5 and 2/6=0.33.
However, stable phases with x $\sim$0.44, 0.61, 0.625, 0.72 and
0.75 have also been observed via phase boundaries studies of solid
state reactions, or by electrochemical
de-intercalation/interaction sweeps.~\cite{Delmas1981,
Fouassier1973}  We note that phases with $x$ equal to an integer
multiple of 1/6 or 1/8 can easily be constructed out of a systemic
loss of Na from a hexagonal basic unit cell (3 Na per unit) or a
rectangular unit cell of $2\sqrt{3}a*2a$ (8 Na per unit). Starting
with fully filled Na-II sites (2/3,1/3,1/4) with $P6_{3}/mmc$
symmetry, the existence of stable phases near x = 1/3 = 0.33 and x
= 2/3 = 0.67 are naturally obtained through systematic loss of Na
from the hexagonal unit. On the other hand, x = 2/8 = 0.25, x =
4/8 = 0.5, x = 5/8 = 0.625 and x = 6/8 = 0.75 can be constructed
through the systematic loss of Na from the $2\sqrt{3}a*2a$ unit.
Most of these proposed superstructures have not been observed with
diffraction methods. It appears likely that only the $x=0.5$
composition has long-range Na order (as reported by Huang $\it{et~
al.}$\cite{Huang2004}).

Our results have also shown that the phases with $x=0.5$ and 0.3
have very similar chemical potentials for Na in the lattice. In
prior work\cite{Chou-mag}, we have observed weak magnetic
susceptibility anomalies near 53K and 88K in some
electrochemically prepared crystals with $x = 0.3$.  This behavior
of the susceptibility is similar to observed in
Na$_{0.5}$CoO$_{2}$ which is reported to have a low temperature
charge- and spin- order state.\cite{Foo2004,Uemura2004} A possible
scenario for the existence of 53K and 88K anomalies in a crystal
with $x=0.3$ is that phase separation of Na occurs over long time
scales. Our current observation of oxygen deficiency in samples
with $x<0.5$ points toward another possibility.  In this scenario,
compounds with $x = 0.5$ will have an average Co valence of +3.5
in the absence of oxygen depletion. If reducing the Na content
simultaneously creates oxygen deficiencies, then the average Co
valence can remain at +3.5, similar to the case of
Li$_{x}$CoO$_{2-\delta}$. ~\cite{Venkatraman2002}  Hence, the weak
anomalies we observe in the susceptibility of
Na$_{0.3}$CoO$_{2-\delta}$ may arise from regions in the sample
with significant oxygen deficiency such that the local average Co
valence remains at +3.5.

In summary, we have presented results characterizing samples of
Na$_{x}$CoO$_2$ prepared by floating-zone crystal growth and
electrochemical de-intercalation.  We have found that the
floating-zone method of crystal growth often yields crystals with
significant amounts of CoO impurity inclusions.  However, the
impurity phase may be minimized by carefully choosing the feeding
rate and pulling rate.  During the de-intercalation process, a
potential step method was employed to identify the stable phases
of Na$_{x}$CoO$_2$.  We find that the stable Na concentrations
correspond to fractions which are close to multiples of 1/6 or
1/8.  This may indicate a tendency for the Na ions to form
short-range ordered superstructures.  Finally, we have observed
that oxygen deficiencies (at a level as high as $\delta \simeq
0.08$) may exist in Na$_{0.75}$CoO$_{2-\delta}$ as well as
de-intercalated Na$_{0.3}$CoO$_{2-\delta}$.  The presence of
oxygen deficiencies must certainly be taken into account when
comparing experimental data on different samples of
Na$_{x}$CoO$_{2-\delta}$.

\begin{acknowledgments}
We thank Patrick Lee and Takashi Imai for many insightful
discussions. This work was supported primarily by the MRSEC
Program of the National Science Foundation under number
DMR-02-13282.  J. H. Cho was partially supported by Grant No.
(R01-2000-000-00029-0) from the Basic Research Program of the
Korea Science and Engineering Foundation.
\end{acknowledgments}

\newpage 
\bibliography{NaxCoO2}

\end{document}